\documentclass[aps, prl, superscriptaddress, amsmath, amssymb, showpacs, twocolumn, notitlepage]{revtex4-1}
\usepackage{amssymb}
\usepackage{amsmath}
\usepackage{graphicx}
\usepackage{bm}
\usepackage{color}
\usepackage{subfigure}
\usepackage{multirow}
\usepackage{colortbl}
\usepackage{color}
\usepackage{soul}
\usepackage{footmisc}

\usepackage[bbgreekl]{mathbbol}
\usepackage{bbm} 
\def\bbm[#1]{\mbox{\boldmath $#1$}}

\graphicspath{{./Figures/}} 

\begin{document}

\title{Giant interatomic energy-transport amplification with nonreciprocal photonic topological insulators}

\author{Pierre Doyeux}
\affiliation{Laboratoire Charles Coulomb (L2C), UMR 5221 CNRS-Universit\'{e} de Montpellier, F- 34095 Montpellier, France}

\author{S. Ali Hassani Gangaraj}
\affiliation{School of Electrical and Computer Engineering, Cornell University, Ithaca, NY 14853, USA}

\author{George W. Hanson}
\affiliation{Department of Electrical Engineering, University of Wisconsin-Milwaukee, 3200 N. Cramer St., Milwaukee, Wisconsin 53211, USA}

\author{Mauro Antezza}
\affiliation{Laboratoire Charles Coulomb (L2C), UMR 5221 CNRS-Universit\'{e} de Montpellier, F- 34095 Montpellier, France}
\affiliation{Institut Universitaire de France, 1 rue Descartes, F-75231 Paris, France}

\begin{abstract}
We show that the energy-transport efficiency in a chain of two-level emitters can be drastically enhanced by the presence of a photonic topological insulator (PTI). This is obtained by exploiting the peculiar properties of its nonreciprocal surface-plasmon-polariton (SPP), which is unidirectional, immune to backscattering and propagates in the bulk bandgap. This amplification of transport efficiency can be as much as two orders of magnitude with respect to reciprocal SPPs. Moreover, we demonstrate that despite the presence of considerable imperfections at the interface of the PTI, the efficiency of the SPP-assisted energy transport is almost unaffected by discontinuities. We also show that the SPP properties allow energy transport over considerably much larger distances than in the reciprocal case, and we point out a particularly simple way to tune the transport. Finally, we analyze the specific case of a two-emitter-chain and unveil the origin of the efficiency amplification. The efficiency amplification and the practical advantages highlighted in this work might be particularly useful in the development of new devices intended to manage energy at the atomic scale.
\end{abstract}

\maketitle

Transporting energy from point A to point B with as little loss as possible is an essential step in countless processes. Focusing on the microscopic scale, one can cite the well-known example of photosynthesis, a natural process in which light is harvested across chromophore complexes from the absorption to the reaction center. On a fundamental level, many efforts have been devoted to unveil the mechanisms at the origin of the efficient energy transport within open quantum systems \cite{plenio_2008, aspuruguzik_2009, kassal_2012, shabani_2014, manzano_2013, gaab_2004, leggio_thermally_2015, doyeux_2017, mostame_2012}.

Besides, the development of new technologies have encouraged engineering of the atomic environment to improve communication between atoms \cite{schachenmayer_2015, poddubny_2015, feist_2015, deroque_2015}, which is of interest for both quantum information theory and energy transport. Among the many possibilities investigated, nonreciprocal systems have been explored \cite{lodahl_2017, pichler_2015, gonzalesballestero_2015b, mahmoodian_2016, kornovan_2017, coles_2016}, taking advantage of the fact that energy exchanges in these systems occur in a privileged direction.

Another successful strategy is to use physical systems where atomic interactions are mediated by surface-plasmon-polaritons (SPPs) \cite{dzsotjan_2010, gonzaleztudela_2011, gonzalezballestero_2015, martincano_2010, andrew_2004}, which are particularly interesting to realize interatomic communication over relatively large distances. However, in such systems, the SPPs usually propagate without a privileged direction, and therefore a significant amount of energy is wasted. Besides, in the presence of imperfections at the interface, the propagation of the SPP can be strongly deteriorated due to scattering, reflection and diffraction, making the practical realizations of such systems sensitive to fabrication errors.

\begin{figure}[h!]
    \centering
    \includegraphics[width = .48 \textwidth]{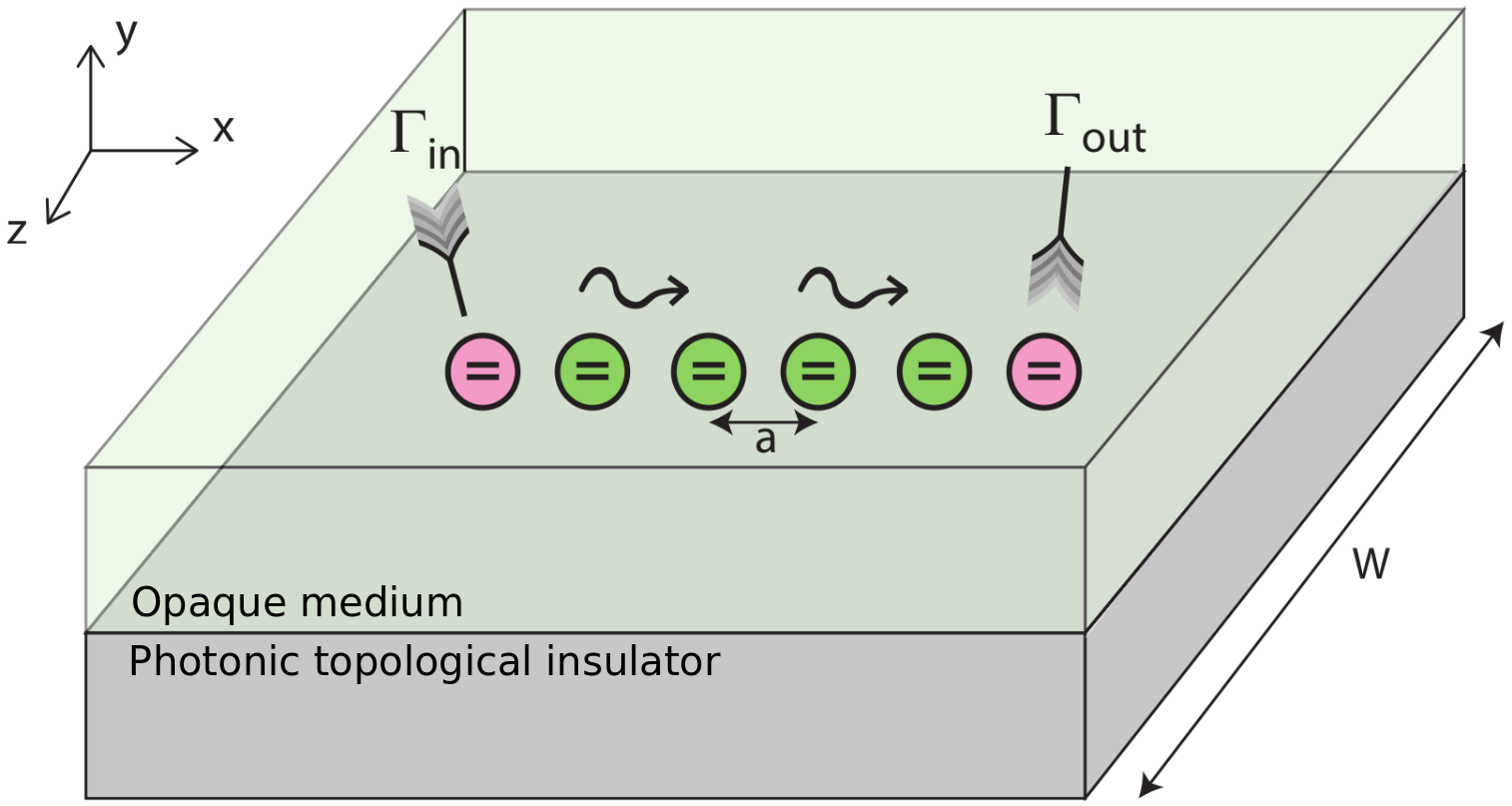}
    \caption{Physical system: chain of $N$ two-level emitters $\{1,2,\ldots,N\}$ located at an interface of width $W$ between a PTI and an opaque medium. The step of the chain is labeled $a$. Depending on the scenario, energy is pumped into atom 1 with rate $\Gamma_\text{in}$ and extracted from atom $N$ with rate $\Gamma_\text{out}$.}
    \label{fig:Physical_system}
\end{figure}

In the past few years, topological insulators have been drawing a lot of attention, principally due to the peculiar behavior of the electronic states occurring at their edges, however, this can seldom be exploited to realize actual one-way transport \cite{torres_2016}. Recently, new physical systems have emerged, the so-called photonic topological insulators (PTIs), showing similar properties but using electromagnetic states. These systems were experimentally observed in photonic crystals \cite{wang_2009}, and theoretically predicted for continuous media \cite{silveirinha_2015}. Remarkably, at the interface of such materials, there can exist unidirectional SPPs that propagate in the bulk bandgap and are immune to backscattering.

\begin{figure}[h!]
    \centering
    \includegraphics[width = .48 \textwidth]{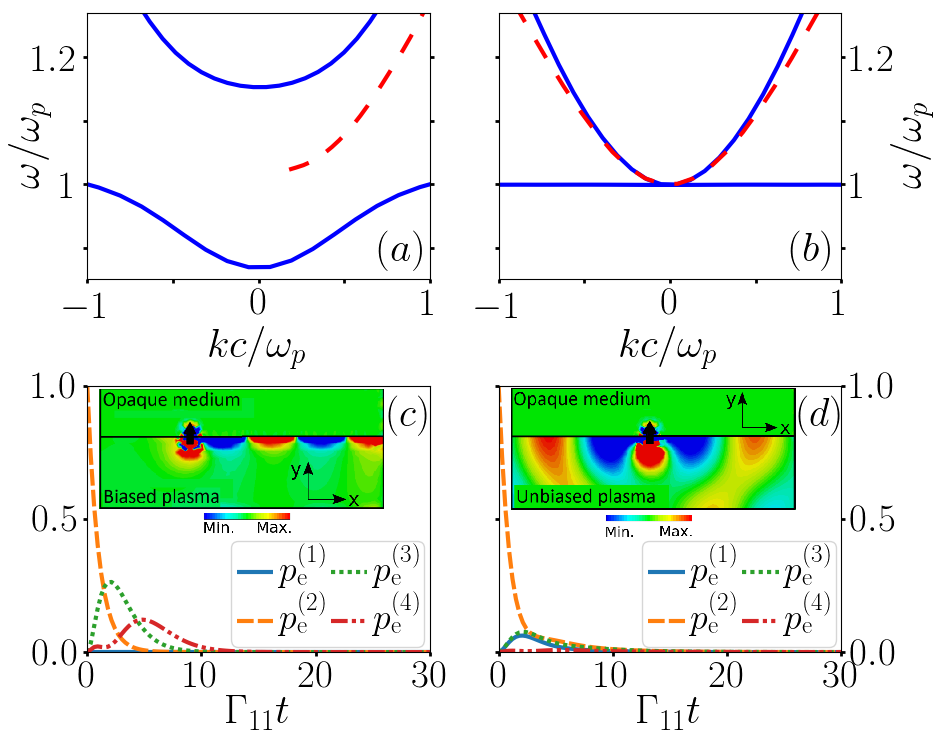}
    \caption{Panels (a)-(b): dispersions of the bulk band (blue solid lines) and SPPs (red dashed line) for different $\bbm[B]$ for interfaces with $W\rightarrow\infty$. The plasma and cyclotron frequencies are denoted $\omega_p$ and $\omega_c\propto\bbm[B]$ \cite{hassani_2017}, respectively, with $\omega_p/\omega_0=0.95$. Panel (a): BP/OM with $\omega_c/\omega_0=0.21$, panel (b): UP/OM ($\omega_c/\omega_0=0$). Panels (c)-(d): dynamics of the excited populations in correspondence with panels (a) and (b), respectively. Inserts : electric field profiles stemming from a single point-source dipole (black arrows).}
    \label{fig:Populations}
\end{figure}

The idea driving this work is to exploit the advantageous properties of the SPP at a PTI interface to produce a significant amplification of the energy-transport efficiency within a chain of two-level emitters (`atoms') with respect to reciprocal interfaces. We highlight the robustness of this amplification against the presence of considerable defects at the interface. Moreover, we show that significant values of efficiency can be reached over a much wider range of interatomic distances with respect to reciprocal environments, and also discuss the possibility of tuning energy transport through an easily accessible parameter. Finally, we focus on the simple case of a two-atom chain to unveil the origin of the efficiency enhancement.\\

\textit{Physical system --} We consider a chain of $N$ two-level atoms with equal transition frequency $\omega_0$. It is supposed that atoms are located at the interface of width $W$ between two different media (Fig.~\ref{fig:Physical_system}) and weakly coupled to their environment. To evaluate energy-transport efficiency, energy will be pumped in atom 1 and extracted from atom $N$ with rates $\Gamma_\text{in}$ and $\Gamma_\text{out}$, respectively.

\begin{figure}[h!]
    \centering
    \includegraphics[width = .48 \textwidth]{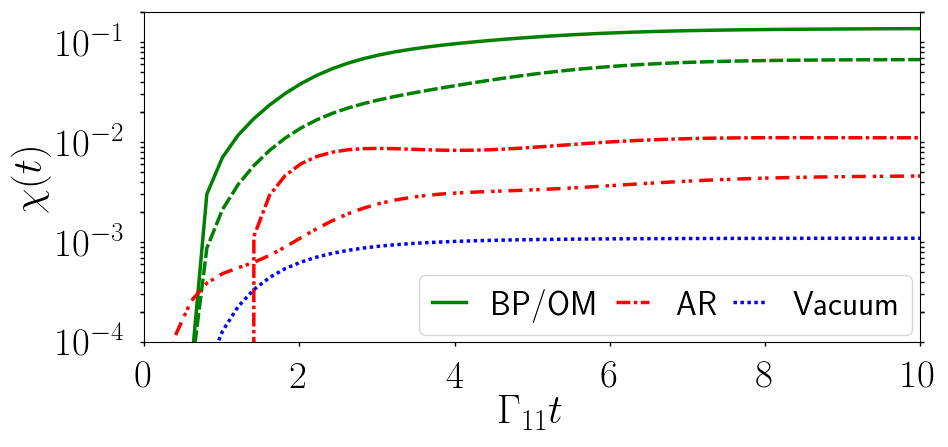}
    \caption{Dynamics of the efficiency of a four-atom chain for several environments. The green solid and red dash-dotted lines correspond to a BP/OM interface of width $W/\lambda_0=1.2$, and the associated AR environment, respectively. Similarly, the green dashed and red dash-double-dotted lines are associated to an infinite-width interface ($W\rightarrow\infty$). The blue dotted line is the vacuum case. The initial state is $|\psi_0\rangle=|g_1g_2g_3e_4\rangle$ and $\Gamma_\text{in}=\Gamma_\text{out}=1.5\,\Gamma_{11}$.}
    \label{fig:Efficiency_R_vs_NR}
\end{figure}

The ground (excited) state of the $i$-th atom is noted by $|g_i\rangle$ ($|e_i\rangle$), and the associated lowering (raising) operator is $\sigma_i=|g_i\rangle\langle e_i|$ ($\sigma_i^\dag=|e_i\rangle\langle g_i|$). The Hamiltonian of the total system reads $H_\text{tot}=H_\text{sys}+H_\text{env}+H_\text{int}$, where $H_\text{sys}=\sum_{i=1}^N \hbar\omega_0\sigma_i^\dag\sigma_i$ ($H_\text{env}$) denotes the bare Hamiltonian of the atomic chain (environment). Under the dipolar approximation, the interaction between the quantum system and its environment is $H_\text{int}=-\sum_{i=1}^N(\sigma_i+\sigma^\dag_i)\bbm[d]_i\cdot\bbm[E](\bbm[r]_i)$, where $\bbm[d]_i$ and $\bbm[r]_i$ are the transition dipole moment and the position of the $i$-th atom, respectively. We will assume that all the atomic dipoles are identical, pointing to the $y$-direction with $|\bbm[d]|=60\,$D. $\bbm[E](\bbm[r]_i)$ denotes the electric field at position $\bbm[r]_i$, which can be expressed in terms of dyadic Green's function, noted $\bbm[G](\bbm[r]_i,\bbm[r]_j,\omega_0)$, describing the response of the medium at point $\bbm[r]_j$ to a point-source dipole located at $\bbm[r]_i$ \cite{hassani_2017,novotny_2006}. In the present work $\bbm[G](\bbm[r]_i,\bbm[r]_j,\omega_0)$ will be computed with finite-element method (COMSOL, \cite{comsol}).

The time evolution of the density matrix associated with the chain is described by the Markovian quantum master equation fully derived in \cite{hassani_2017} using the Born-Markov and rotating-wave approximations. The master equation valid for reciprocal and nonreciprocal environments is \cite{hassani_2017}:
\begin{multline}
    \dot{\rho}(t)=-\frac{i}{\hbar}\big[\text{H}_\text{sys},\rho(t)\big]
    +\bigg\{\sum_{i=1}^{N}\frac{\Gamma_{ii}}{2}\mathcal{D}(\sigma_i)\\
   +\sum_{i\neq j}\Big(\frac{\Gamma_{ij}}{2}\tilde{\mathcal{D}}(\sigma_i,\sigma_j)+g_{ij}\tilde{\mathcal{D}}(i\sigma_i,\sigma_j)\Big)\\
    +\frac{\Gamma_\text{in}}{2}\mathcal{D}(\sigma_1^\dag)+\frac{\Gamma_\text{out}}{2}\mathcal{D}(\sigma_N)\bigg\}[\rho(t)],
    \label{eq:master_equation}
\end{multline}
where the following super-operators have been introduced: $\mathcal{D}(\sigma_i)[\rho(t)]=2\sigma_i\rho(t)\sigma_i^\dag-\sigma_i^\dag\sigma_i\rho(t)-\rho(t)\sigma_i^\dag\sigma_i$ and $\tilde{\mathcal{D}}(\sigma_i,\sigma_j)[\rho(t)]=[\sigma_j\rho(t),\sigma_i^\dag]+[\sigma_i,\rho(t)\sigma_j^\dag]$.  The coefficients $g_{ij}$ and $\Gamma_{ij}$ are the coherent and dissipative rates, respectively, which depend on the Green's function as
\begin{align}
    \Gamma_{ij}=\frac{2\omega_0^2}{\varepsilon_0\hbar c^2} \text{Im}\big[\bbm[d]\cdot\bbm[G](\bbm[r]_i,\bbm[r]_j,\omega_0)\cdot\bbm[d]\big], \\
    g_{ij}=\frac{\omega_0^2}{\varepsilon_0\hbar c^2} \text{Re}\big[\bbm[d]\cdot\bbm[G](\bbm[r]_i,\bbm[r]_j,\omega_0)\cdot\bbm[d]\big].
    \label{eq:ME_coeff}
\end{align}

\begin{figure}[h!]
    \centering
    \includegraphics[width = .48 \textwidth]{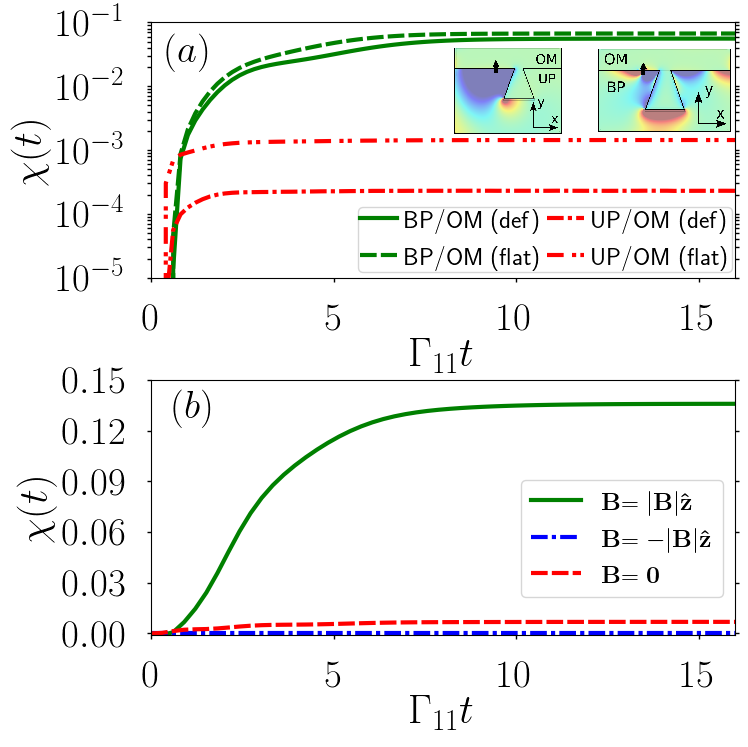}
    \caption{Panel (a): Efficiency dynamics of a four-atom chain for interfaces of width $W\rightarrow \infty$, where `def' (`flat') stands for the presence (absence) of defect between the atoms 2 and 3. The length of the defect contour is $\sim1.8\,\lambda_0$. The electric field profiles in the presence of defect are also shown. Panel (b): Efficiency dynamics depending on $\bbm[B]$ for an interface of width $W/\lambda_0=1.2$.}
    \label{fig:Robustness_and_control}
\end{figure}

At the interface (Fig.~\ref{fig:Physical_system}) between a topologically nontrivial (with Chern number $C=1$) biased plasma (BP), i.e. a plasma biased by a magnetic field $\bbm[B]=|\bbm[B]|\bbm[\hat{z}]$ \cite{yang_2016}, and an opaque medium (OM) with permittiviy $\varepsilon=-2$ ($C=0$), there exists a topologically-protected backscattering-immune and unidirectional SPP that spans the common bulk bandgap (Fig.~\ref{fig:Populations}(a)). {The nonreciprocal character implied by the permittivity of the biasable plasma we consider here has been studied in \cite{seshadri_1962}.} Assuming this SPP propagates from $\bbm[r]_i$ to $\bbm[r]_j$, then $\bbm[G](\bbm[r]_i,\bbm[r]_j,\omega_0)\neq0$ and $\bbm[G](\bbm[r]_j,\bbm[r]_i,\omega_0)=0$.

When $\bbm[B]=\bbm[0]$, the topology of the plasma is trivial ($C=0$), leading to a reciprocal SPP at the interface of unbiased plasma--opaque medium (UP/OM) (Fig.~\ref{fig:Populations}(b)), such that $\bbm[G](\bbm[r]_i,\bbm[r]_j,\omega_0)=\bbm[G](\bbm[r]_j,\bbm[r]_i,\omega_0)$.

Regarding energy transport, unidirectional environments seem clearly more advantageous compared to omnidirectional environments, since energy can only move in one direction. To illustrate this in a simple case, we set a chain of step $a/\lambda_0=0.9$ with $N=4$ and $\Gamma_\text{in}=\Gamma_\text{out}=0$ (no pumping nor extraction). Figures~\ref{fig:Populations}(c) and \ref{fig:Populations}(d) show the excited populations of this chain defined as $p_e^{(i)}(t)=\text{Tr}(\sigma_i^\dag\sigma_i\rho(t))$ with $i=1,\ldots,4$. The initial state $|\psi_0\rangle=|g_1e_2g_3g_4\rangle$ is such that only atom 2 is excited whereas the others are in their ground states. At the BP/OM interface, the initial excitation travels along the chain from atom 2 to atom 4, whose maximum probability of excitation reaches a significant value. Remarkably, atom 1 remains strictly in its ground state throughout the evolution, highlighting the unidirectionality of the SPP assisting energy exchanges ($g_{j1}=\Gamma_{j1}=0$, for $j\in\{2,3,4\}$).

In the case of the UP/OM interface, Fig.~\ref{fig:Populations}(d), the SPP is reciprocal, which explains why the initial excitation of atom 2 is transmitted to its neighbors on both sides. In particular, atom 1 is affected by the presence of the excitation, contrary to the unidirectional case. Furthermore, the excitation is lost to the bulk regions before being able to reach the atom 4, whose probability of excitation remains negligible.\\

\textit{Efficiency amplification in atomic chains --} {To evaluate the energy-transport efficiency, we solve two different master equations. The no-pumping (pumping) scenario is characterized by $\Gamma_\text{in}=0$ and $\Gamma_\text{out}\neq0$ ($\Gamma_\text{in},\Gamma_\text{out}\neq0$), and the corresponding solution is noted $\rho_0$ ($\rho$). Introducing, for a generic solution $\tilde{\rho}(t)$, the energy fluxes of pumping $P(\tilde{\rho}(t))=\dfrac{\Gamma_\text{in}}{2}\text{Tr}\Big(H_\text{sys}\mathcal{D}(\sigma_1^\dag)[\tilde{\rho}(t)]\Big)$ and extraction $E(\tilde{\rho}(t))=-\dfrac{\Gamma_\text{out}}{2}\text{Tr}\Big(H_\text{sys}\mathcal{D}(\sigma_N)[\tilde{\rho}(t)]\Big)$, our definition of transport efficiency reads
\begin{equation}
    \chi(t)=\frac{E(\rho(t))-E(\rho_0(t))}{P(\rho(t))}.
    \label{eq:efficiency}
\end{equation}
}
When no additional energy is extracted despite the pumping then $\chi(t)=0$, whereas $\chi(t)=1$ indicates that the pumped energy is transported along the chain without any loss.

In the following, we will use Eq.~\eqref{eq:efficiency} to compare transport efficiency between reciprocal and unidirectional environments in several situations. According to the properties of $\chi(t)$, we have to distinguish two different reciprocal environments. It is natural to compare UP/OM and BP/OM interfaces, since the difference depends only on the absence or presence of the biasing field. However, the SPPs existing in these two situations have intensities of different order of magnitude. To unveil the effect of one-wayness on $\chi(t)$, we have to compare systems with SPPs having the same properties (excitation amplitude, confinement factor, etc.). Thus, starting from a unidirectional environment, we set $\Gamma_{ij}=\Gamma_{ji}$ and $g_{ij}=g_{ji}$, and multiply the rates $\Gamma_{ii}$ by 2 \cite{gonzalesballestero_2015b}, so that we have an artificially-reciprocal (AR) medium, comparable to the biased plasma case.

Figure \ref{fig:Efficiency_R_vs_NR} shows the dynamics of $\chi(t)$ of a four-atom chain in different environments. More specifically, one can compare it between a finite-width ($W<\infty$) BP/OM interface with the corresponding AR environment. The nonreciprocal environment produces a stationary efficiency $\chi(\infty)$ much better than in the reciprocal case, with a considerable amplification of 1 order of magnitude, passing from a negligible efficiency of $1\%$ to a significantly improved and significant value of $15\%$. This amplification is even larger for an infinite-width ($W\rightarrow\infty$) BP/OM interface and its associated AR environment.

Although the values of efficiency considered here might seem relatively low, optimizing $\chi(t)$ with respect to all the parameters of the system could unveil configurations with much higher efficiency. This, however, is beyond the scope of this work, {whose main purpose is to highlight the possibility of amplifying significantly $\chi$ using PTIs. Moreover, another important aspect of PTIs regarding energy transport lies in the practical advantages they offer, three of them being presented below.}

{Firstly}, when operated in the bulk bandgap, radiation is suppressed into the bulk, and is focused into the SPP, even in the presence of surface discontinuities. Thus, in the presence of defect, the unidirectional and backscattering-immune SPP bypasses the obstacle. In Fig.~\ref{fig:Robustness_and_control}(a), {an electrically-large defect is introduced by considering a trapezoidal-shaped deformation of the flat interface (see the insets)}, and clearly $\chi(t)$ is hardly affected in the nonreciprocal environment, while it is strongly diminished in the reciprocal one. In this case, the efficiency is amplified by more than 2 orders of magnitude between UP/OM and BP/OM interfaces.

{Secondly, PTIs also offer the capacity of changing $\chi$ by modifying the orientation of the biasing field \bbm[B]. Indeed, this operation, which has the advantage of being easily achievable, amounts to change the direction of propagation of the unidirectional SPP. This is illustrated with Fig.~\ref{fig:Robustness_and_control}(b): the green solid line has been obtained with $\bbm[B]=|\bbm[B]|\bbm[\hat{z}]$, and is such that the energy pumped in the first atom is (partially) transported along the chain. On the other hand, reverting the direction of the field such that $\bbm[B]=-|\bbm[B]|\bbm[\hat{z}]$ (blue dashed line) induces that this energy can only be dissipated into the environment, thus leading to $\chi(t)=0$. Having $\bbm[B]=\bbm[0]$ results in a reciprocal SPP with an intensity much lower than in the biased case, leading to a negligible efficiency.}

\begin{figure}[h!]
    \centering
    \includegraphics[width = .48 \textwidth]{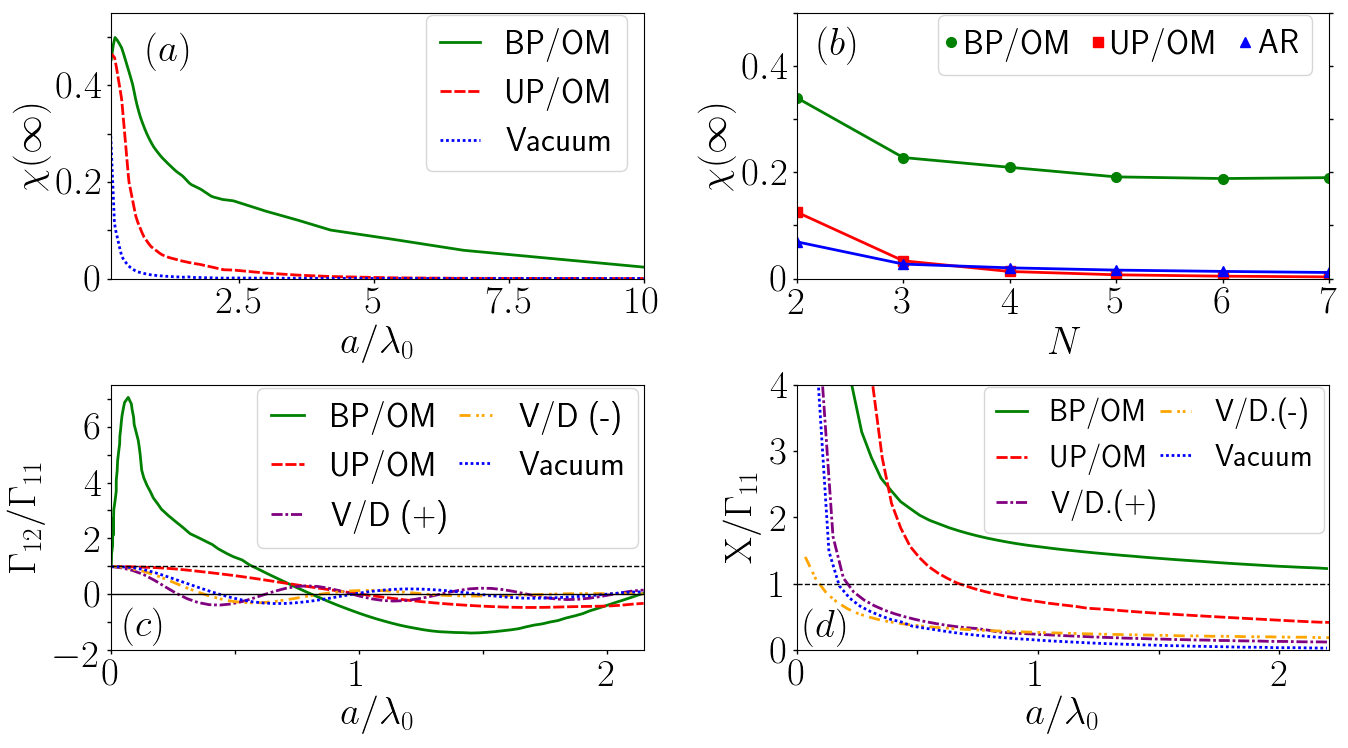}
    \caption{Panel (a): Efficiency as a function of the chain step ($N=2$). Panel (b): Efficiency against the number of atoms composing the chain, with $a/\lambda_0=0.6$. In both panels $(a)$ and $(b)$, the interface width is $W/\lambda_0=1.2$. Panels (c) and (d) share the same legend, where V/D(+) ((V/D(-)) indicates an interface vacuum--dielectric with permittivity $\varepsilon=2$ ($\varepsilon=-2$). Panel (c): Collective dissipative rate against the interatomic separation. Panel (d):  Green's function modulus for the interfaces in correspondence with panel (c). In the unidirectional case, only $X$ contributes to the efficiency. }
    \label{fig:Mixe}
\end{figure}

{Thirdly}, another effect of nonreciprocity is to increase significantly the range of energy transport. Figure~\ref{fig:Mixe}(a) shows the stationary efficiency of a two-atom chain as a function of $a/\lambda_0$. Clearly, $\chi(\infty)$ survives over distances much greater when the environment is nonreciprocal rather than reciprocal. For instance, having $\chi(\infty)=0.1$ with the UP/OM interface necessitates $a\sim0.7\,\lambda_0$, while the same value is reached for a distance $\sim6\times$ greater for the BP/OM interface.

{Finally}, Fig.~\ref{fig:Mixe}(b) shows $\chi(\infty)$ as a function of $N$ for the nonreciprocal and the two reciprocal environments. Not only is the efficiency much better with the BP/OM interface, but also it remains almost constant despite the increase of the number of atoms, in contrast with the two reciprocal environments.\\

\textit{Physical insight with two atoms --} To unveil the origin of the efficiency amplification, we focus on the energy-transport of a two-atom chain.

We make the assumption that the main contribution to the Green's function at the interface comes from the SPP. There is a quadrature relation between the real and imaginary parts of the SPP Green function \cite{martincano_2011}. Thus, the coefficients describing the different energy channels are
\begin{align}
    g_{ij}&=\frac{1}{2}X\cos(\phi), \,\,\,i\neq j,\\
    \Gamma_{ij}&=X\sin(\phi), \,\,\,i\neq j,\\
    \Gamma_{11}&=\Gamma_{22},
\end{align}
where $i,j\in\{1,2\}$. The parameters $X$ and $\phi$ both depend on the atomic positions such that $\lim\limits_{a\rightarrow0}\Gamma_{ij}=\Gamma_{ii}$, where $a=|\bbm[r]_j-\bbm[r]_i|$. In the absence of biasing field, the environment is reciprocal, i.e. $\Gamma_{21}=\Gamma_{12}$ and $g_{21}=g_{12}$, while in the unidirectional case $\Gamma_{21}=g_{21}=0$.

In the following, our aim is to determine what is the best environment between reciprocal and unidirectional in terms of $\chi(\infty)$. Therefore, we have to determine the appropriate values of $X$ and $\phi$ that produce the best efficiency in each environment.

Numerical simulations (not shown here) show that the stationary efficiency for a fixed $X$ does not depend on $\phi$ in the unidirectional case (except in the limit $a\rightarrow0$), while it is maximum when $\phi=\pi/2$ in the reciprocal one. Thus, we set $\phi=\pi/2$ hereafter.

We now have to determine the value of $\Gamma_{12}$ optimizing the efficiency in each case. Figure~\ref{fig:Mixe}(c) shows the ratio $\Gamma_{12}/\Gamma_{11}$ as a function of $a/\lambda_0$ for several realistic interfaces. In all of the reciprocal environments, this ratio is $\leq1$, suggesting that $\Gamma_{12}$ is bounded by $\Gamma_{11}$. As a more general argument, in order to have a valid reciprocal master equation, the matrix associated to the dissipative rates must be positive \cite{breuer_theory_2002}. In the case of two identical atoms, this condition is verified precisely when $\Gamma_{12}\leq\Gamma_{11}$.

\begin{figure}[h!]
    \centering
    \includegraphics[width = .48 \textwidth]{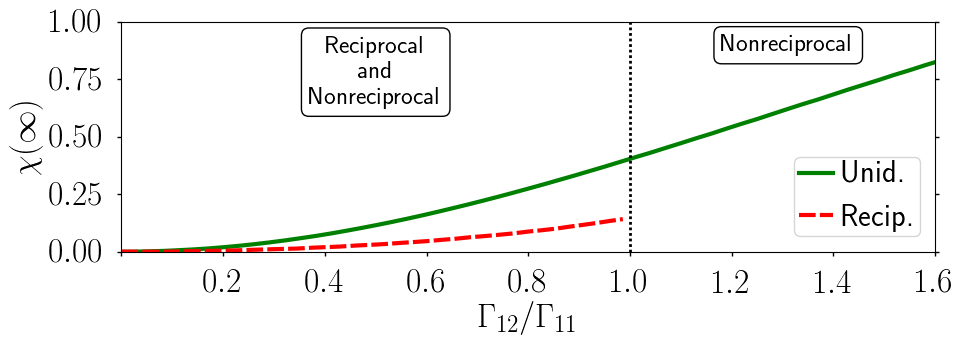}
    \caption{Stationary efficiency as a function of the rate $\Gamma_{12}$ for $\phi=\pi/2$. The unidirectional case has been obtained artificially from the reciprocal one.}
    \label{fig:Efficiency_vs_Xnormalized}
\end{figure}

The structure of the master equation is different for nonreciprocal environments, and the condition $\Gamma_{12}\leq\Gamma_{11}$ does not necessarily apply, e.g. with the BP/OM interface displayed in Fig.~\ref{fig:Mixe}(c) (green solid line), where $\Gamma_{12}>\Gamma_{11}$ for many atom spacings. Consequently, the atoms-SPP coupling is considerably stronger than in the reciprocal case, leading to a better efficiency.

Figure \ref{fig:Efficiency_vs_Xnormalized} represents $\chi(\infty)$ in reciprocal environment (red dashed line) when $\phi=\pi/2$ as a function of $\Gamma_{12}/\Gamma_{11}$ in the range $[0,1]$. In the same spirit as for Fig.~\ref{fig:Efficiency_R_vs_NR}, the efficiency for the unidirectional environment in Fig.~\ref{fig:Efficiency_vs_Xnormalized} (green solid line) has been obtained artificially starting from the reciprocal one, by setting $\Gamma_{21}=0$ and dividing the coefficient $\Gamma_{11}$ by 2.

This plot is particularly revealing: firstly, it shows that unidirectional environments always produce a better efficiency in the range $\Gamma_{12}/\Gamma_{11}\in[0,1]$. Secondly, $\chi(\infty)$ reaches values even higher in a region of parameters forbidden to reciprocal environments. Thirdly, the best efficiency in reciprocal environments necessitates having $\Gamma_{12}=\Gamma_{11}$, which is obtained only in the limit $a\rightarrow0$. On the contrary, the configuration $\Gamma_{12}>\Gamma_{11}$ is easily accessible in nonreciprocal environments for a wide range of values of $a$.

\textit{Conclusion --} {We have investigated energy transport within an $N$-atom chain nearby a PTI, an emerging material allowing the existence of one-way SPPs at its surface. Exploiting this property, we have shown that transport efficiency can be drastically enhanced by 1 order of magnitude with respect to comparable reciprocal environments.}

{We have also highlighted relevant practical advantages regarding energy transport stemming from these new materials. In particular, we have pointed out the remarkable robustness of the efficiency against the presence of discontinuities at the interface, in which case the amplification, being of 2 orders of magnitude, is even more striking.  Moreover, we have demonstrated the possibility of tuning energy transport by modifying the biasing field, which additionally has the advantage of being an easily accessible parameter. We have also shown that energy can be transported over a much larger range than reciprocal environments, and that adding atoms to the chain still produces a significant efficiency.}

{More fundamentally, we have analyzed the case of a two-atom chain and demonstrated that the efficiency amplification stems from a stronger atoms-SPP coupling, unattainable in reciprocal environments.}

{All the aspects presented here suggest that the PTIs are promising candidates regarding further development of emerging technologies requiring efficient and tunable energy-transport at the microscopic scale, such as quantum technologies and energy management.}

\bibliographystyle{apsrev4-1}
\bibliography{References}

\end{document}